\documentclass[12pt]{article}

\usepackage{amssymb}
\usepackage{amsmath}  
\usepackage{amsbsy}
\numberwithin{equation}{section}

\newcommand{\dis}{\displaystyle}
\newcommand{\dvp}{\dot{\vec{\phi}\;}\!}

\baselineskip=\normalbaselineskip

\begin{document}

{\par\raggedleft PUPT-1917\\
hep-th/0002106\par}

{\par\centering \textbf{\LARGE Gauge fields}{\LARGE \,--\,}\textbf{\LARGE strings
duality and the loop} {\LARGE }\textbf{\LARGE equation}{\LARGE \vspace{1cm}}\LARGE \par}

{\par\centering {\large A. Polyakov\( ^{1} \) and V. Rychkov\( ^{2} \)\vspace{1cm}}\large \par}

{\par\centering \( ^{1} \) Joseph Henry Laboratories, Princeton University,
Princeton, NJ 08544\par}

{\par\centering polyakov@viper.princeton.edu\par}

{\par\centering \( ^{2} \) Department of Mathematics, Princeton University,
Princeton, NJ 08544\par}

{\par\centering rytchkov@math.princeton.edu\par}

\section*{\centerline{Abstract}}

\noindent We explore gauge fields\,--\,strings duality by means of the loop
equations and the zigzag symmetry. The results are striking and incomplete.
Striking---because we find that the string ansatz proposed in \cite{1} satisfies
gauge theory Schwinger-Dyson equations precisely at the critical dimension \( D_{\textrm{cr}}=4 \).
Incomplete---since we get these results only in the WKB approximation and only
for a special class of contours. The ways to go beyond these limitations and
in particular the OPE for operators defined on the loop are also discussed.
\vspace{1cm}

\noindent February 2000

\pagebreak

\section{Introduction}

Gauge fields\,--\,strings duality is an old and fundamental subject. It underwent
a rapid and fascinating development in the last three years. In this duality
color-electric flux lines of gauge theory are described as certain relativistic
strings. It has been shown in \cite{1} that the ``natural habitat'' for these
strings (and thus for the flux lines) is a five-dimensional curved space with
the metric
\begin{equation}
\label{1.1}
ds^{2}=d\phi ^{2}+a^{2}(\phi )\, d\vec{x}\, ^{2}\; .
\end{equation}
The function \( a^{2}(\phi ) \) (which represents the running string tension)
must be of a special type. In order to properly describe the zigzag-invariant
Wilson loop, it must have a horizon, where \( a^{2}(\phi _{H})=0 \), and an
infinity where \( a^{2}(\phi _{I})=\infty  \). The precise form of \( a^{2}(\phi ) \)
is determined from the condition of conformal invariance on the world sheet.
The gauge fields\,--\,strings duality was formulated in \cite{1} as an isomorphism
between the closed string vertex operators and the gauge invariant operators
of gauge theory.

Another development \cite{2} was related to the \( \mathcal{N}=4 \) SYM theory
and the 3-branes in the type IIB strings. It has been shown there that some
of the SYM correlation functions (describing absorption of the external particles
by the 3-brane) can be calculated by the use of the classical supergravity in
which the 3-brane is replaced by the metric (\ref{1.1}). It was further conjectured
\cite{3} that the correspondence extends beyond the SUGRA approximation and
the function \( a^{2}(\phi ) \) in this case must correspond to the AdS space
(\( a^{2}(\phi )\varpropto e^{\alpha \phi } \)).

In the subsequent papers \cite{4},\( \,  \)\cite{5} it was shown how to implement
the above-mentioned isomorphism in the \( \mathcal{N}=4 \) SYM theory. After
that the correspondence in this case has been confirmed in an almost infinite
number of papers.

But all is not well. There is no real understanding (beyond the heuristic arguments
of \cite{1}-\cite{5}) why and, more importantly, when the correspondence works.
Ideally, one would like to check that the Schwinger-Dyson equations of the Yang-Mills
theory can be obtained from the string representation. This would clarify which
string theories must be used for the various gauge theories. It is important
to stress in this respect that the help of the D-branes in answering the above
question is not always available. There are reasons to believe that in the most
interesting non-supersymmetric cases the D-brane interpretation of the general
\( \sigma  \)-model metric (\ref{1.1}) is not possible. In these cases the
Schwinger-Dyson equations (formulated as loop equations) is our only tool.

In the vast literature on the subject there have been occasional attempts to
explore the loop equations, but no conclusive results have been reached so far.

The reason, we believe, lies in some common misconceptions concerning loop equations.
It is generally thought that the loop operator is singular and can be applied
only to the regularized and non-universal Wilson loop. This is not necessarily
so. In \cite{6} it has already been sketched how to implement the loop equation
for the renormalized loops.

In the present paper we shall apply the loop operator to the string functional
integral. The results are incomplete but quite stunning. We will be dealing
only with the very special contours---wavy lines (suggested in \cite{1}) and
only in the WKB approximation. This is obviously incomplete. We will find in
this case that, first, the loop operator is well defined in any dimension (implying
the zigzag symmetry of the string representation), and second, at the critical
dimension \( D_{\textrm{cr}}=4 \) the loop equation is satisfied. This we find
quite stunning.

\section{The loop equation and the zigzag symmetry}

The loop equation has been introduced in \cite{7} and further explored in \cite{6}
and \cite{8} (see also some related developments in \cite{9},\( \,  \)\cite{10}).
We will give its derivation now, which can be used for the renormalized Wilson
loop. It will contain some additional elements to the old works. The Wilson
loop is given by
\begin{equation}
\label{2.1}
W[C]=\frac{1}{N}\left\langle \textrm{Tr}P\exp \oint _{C}A_{\mu }dx^{\mu }\right\rangle \; .
\end{equation}
 The averaging in this formula is performed with the Yang-Mills action
\[
S=\frac{1}{4g_{YM}^{2}}\int \textrm{Tr}F_{\mu \nu }^{2}\, (dx)\]
 (\( F_{\mu \nu } \) is the Yang-Mills field strength). The idea of the loop
equation is to find an operation in the loop space which, being applied to the
LHS of (\ref{2.1}), will give the Yang-Mills equations of motion at the RHS.
To implement this idea, consider the second variational derivative of \( W[C] \)
\begin{eqnarray}
\frac{\delta ^{2}W}{\delta x_{\mu }(s)\delta x_{\mu }(s')}=\left\langle \textrm{Tr}\, P\left( \nabla _{\mu }F_{\mu \nu }(x(s))e^{\oint A_{\mu }dx_{\mu }}\right) \right\rangle \dot{x}_{\nu }(s)\delta (s-s')+ &  & \nonumber \\
\left\langle \textrm{Tr}\, P\left( F_{\mu \lambda }(x(s))F_{\mu \sigma }(x(s'))e^{\oint A_{\mu }dx_{\mu }}\right) \right\rangle \dot{x}_{\lambda }(s)\dot{x}_{\sigma }(s') &  & \label{2.2} 
\end{eqnarray}
 (where \( P \) is the ordering along the contour).

The main idea of the loop equation is to separate the first term, and to introduce
the ``loop Laplacian'' which acting on \( W[C] \) gives the Yang-Mills equation
of motion. This can be achieved in several different ways. The problem to overcome
is the singularity of the second term at \( s=s' \), which makes it difficult
to distinguish it from the first one.

The ``brute force'' method would be to regularize the gauge theory with some
cut-off \( \Lambda  \) and the consider \( |s-s'|\ll 1/\Lambda  \). The second
term is regular in this case, while the first one contains the \( \delta  \)-function
which is easy to pick up. The price to pay is the non-universal arbitrary regularization.
We cannot be sure that on the string side the regularization is the same and
thus the comparison of gauge fields and strings becomes strictly speaking impossible.
It has already been noticed in \cite{6} that the way out of this difficult
is to use the operator product expansion as \( s\to s' \). 

Consider a set of operators \( \left\{ \mathcal{O}_{n}(x)\right\}  \) which
are \emph{not} color singlets. We can in general examine a gauge invariant amplitude
\[
G_{n_{1}\ldots n_{N}}(s_{1},\ldots s_{N})=\left\langle \textrm{Tr}P\left( \mathcal{O}_{n_{1}}(x(s_{1}))\mathcal{O}_{n_{2}}(x(s_{2}))\ldots e^{\oint _{C}A_{\mu }dx_{\mu }}\right) \right\rangle \; .\]
 Let us assume that the contour \( C \) is smooth and non-selfintersecting.
In this case we expect the OPE to have the form\footnote{%
Such operator products were considered before in \cite{16}.
}
\begin{equation}
\label{2.4}
\mathcal{O}_{n_{1}}(x(s_{1}))\mathcal{O}_{n_{2}}(x(s_{2}))=\sum \frac{C^{m}_{n_{1}n_{2}}(x(s))}{\left| x(s_{1})-x(s_{2})\right| ^{\Delta _{n_{1}}+\Delta _{n_{2}}-\Delta _{m}}}\mathcal{O}_{m}(x(s))
\end{equation}
 (where \( s=\frac{s_{1}+s_{2}}{2} \) and \( \Delta _{n} \) are anomalous
dimensions of the corresponding operators). Of course, in asymptotically free
theories there are also powers of logarithms \( \log \left| x(s_{1})-x(s_{2})\right|  \)
in these formulas, which we do not display.

In this way the OPE in the physical 4d space is transplanted to the one-dimensional
contour \( C \). The ``open string'' amplitudes \( G_{n_{1}\ldots n_{N}}(s_{1},\ldots s_{N}) \)
have power singularities at the coinciding points. There is no room for the
contact terms proportional to \( \delta (s-s') \) if we use renormalized correlators.
Moreover, the possible contact terms in the \( x \)-space do not give contact
terms in the \( s \)-space. Consider as an example a singularity
\[
\delta (x(s)-x(s'))=\lim _{\Delta \to 4}(4-\Delta )\frac{1}{|x(s)-x(s')|^{\Delta }}\; .\]
 As it was explained in \cite{6}, when one uses test functions,
\[
\int \delta (x(s)-x(s'))\, f(s')\, ds'\, =\lim _{\Delta \to 4}(4-\Delta )\int \frac{f(s')\, ds'}{|x(s)-x(s')|^{\Delta }}=0\; ,\]
 provided that the contour has no self-intersections, so that the only singularity
in the integral comes from \( s=s' \). It is also assumed that the integral
is defined by analytic continuation.

As a result of this discussion we conclude that the second variational derivative
of the renormalized (and analytically regularized) Wilson loop has the form
\begin{equation}
\label{2.6}
\frac{\delta ^{2}W}{\delta x_{\mu }(s)\delta x_{\mu }(s')}=(\widehat{{L}}(s)W)\delta (s-s')+\sum _{n}\frac{C_{n}(s,\{x(s)\})}{|x(s)-x(s')|^{4-\Delta _{n}}}\; ,
\end{equation}
 where \( C_{n}(s,\{x(s)\})\varpropto \left\langle \textrm{Tr}P\left( \mathcal{O}_{n}(x(s))e^{\oint A\, dx}\right) \right\rangle  \).
Since \( |x(s)-x(s')|\approx \sqrt{\dot{x}^{2}(s)}|s-s'| \), the nonlocal power-like
behavior of the second term for \( s\to s' \) can be separated from the local
\( s=s' \) singularity of the first one. The loop Laplacian is defined as the
coefficient \( \widehat{{L}}(s)W \) in front of the \( \delta  \)-function,
and the loop equation for non-selfintersecting contours has the form
\begin{equation}
\label{2.65}
\widehat{L}(s)W=0\; .
\end{equation}

In our applications it will be convenient to write (\ref{2.6}) in the momentum
space:
\begin{equation}
\label{2.7}
\lim _{p\to \infty }\frac{\delta ^{2}W}{\delta x_{\mu }\left( \frac{q}{2}+p\right) \delta x_{\mu }\left( \frac{q}{2}-p\right) }=(\widehat{{L}}_{q}W)p^{0}+\sum _{n}C_{n}(q)|p|^{\lambda _{n}}\; .
\end{equation}
 This is an important formula. It shows that the Wilson loop belongs to a very
special class of functionals, for which the asymptotics in (\ref{2.7}) does
not contain integer even powers of \( p \), corresponding to the derivatives
of the \( \delta  \)-function. If it does, this means that the functional \( W[C] \)
\emph{cannot} be represented as an ordered exponential (\ref{2.1}). 

Some danger for this analysis would be presented by odd integers \( \Delta _{n} \)
in (\ref{2.6}), which would give a \( p^{k}(\log p+\textrm{const}) \) contribution.
However, the lowest operators which can appear in (\ref{2.4}), even when protected
by non-renormalization theorem, have even dimensions. An exception from this
is the operator \( \nabla _{\alpha }F_{\beta \gamma } \) which can appear in
the OPE:
\[
F_{\mu \lambda }(x(s))F_{\mu \sigma }(x(s'))\sim \frac{(x(s)-x(s'))_{\mu }\nabla _{\mu }F_{\lambda \sigma }}{\left| x(s)-x(s')\right| ^{2}}\; .\]
 (We assume here the scenario in which the \( F_{\mu \nu } \) operator has
the normal dimension as a result of some non-renormalization theorem.) However,
the singular contribution from this term to (\ref{2.2}) vanishes after contraction
with \( \dot{x}_{\lambda }(s)\dot{x}_{\sigma }(s') \). 

The above analysis can be extended to the functionals represented by the iterated
integrals 
\begin{equation}
\label{2.8}
W[C]=\sum _{n}\int _{s_{1}<s_{2}\ldots <s_{n}}\Gamma ^{(n)}_{\mu _{1}\ldots \mu _{n}}(x(s_{1}),\ldots x(s_{n}))\, \dot{x}_{\mu _{1}}(s_{1})\ldots \dot{x}_{\mu _{n}}(s_{n})\, ds_{1}\ldots ds_{n}
\end{equation}
 (of which (\ref{2.1}) is a special case). Such functionals behave well under
the application of the loop Laplacian (see \cite{8}). Hence, the second variational
derivative for them has the structure (\ref{2.7}).

Another feature of the functionals (\ref{2.8}) is the zigzag symmetry \cite{1}.
When we change \( s\Rightarrow \alpha (s) \), they are invariant even if \( \alpha '(s) \)
changes sign (and is not a diffeomorphism). It is likely that any zigzag-invariant
functional can be represented in the form (\ref{2.8}), if it satisfies some
extra analyticity requirements. Without them many other forms are possible,
e.g.
\[
I_{\alpha \beta }=\int ds\left( \frac{\ddot{x}_{\alpha }\dot{x}_{\beta }-\ddot{x}_{\beta }\dot{x}_{\alpha }}{\dot{x}^{2}}\right) \, .\]
 At present the precise form of these requirements is not known, although a
possibility is that the functional must be nonsingular with respect to \( \dot{x}^{2} \).
In the following, we will use the term ``zigzag symmetry'' referring to the
form (\ref{2.8}). According to the above discussion, there is a \emph{practical
way} for testing the zigzag symmetry in this sense: take the asymptotics (\ref{2.7})
and check for the \emph{}terms \emph{\( \varpropto p^{n} \)} with \( n \)---positive
even integer. If they are present, there is no place for the zigzag symmetry,
and the functional does not represent any Yang-Mills theory. This conclusion,
together with the formula (\ref{2.7}), will be extensively used below.

\section{The string representation: D-branes and/or \protect\( \sigma \protect \)-model.}

Let us summarize the facts about string representation of the Wilson loop. According
to \cite{1} we must consider a 5d background with the metric (\ref{1.1}).
The origin of the fifth dimension is the Liouville degree of freedom which is
unavoidable in the non-critical strings. The \( \sigma  \)-model action of
this string is given by
\begin{eqnarray}
S & = & \frac{1}{2}\int d^{2}\xi \sqrt{{g}}g^{ab}(\xi )G_{MN}(z(\xi ))\partial _{a}z^{M}\partial _{b}z^{N}+\Phi (z(\xi ))\, ^{(2)}\! R(g)\sqrt{{g}}\nonumber \\
 &  & \quad +\, \epsilon ^{ab}B_{MN}(z(\xi ))\partial _{a}z^{M}\partial _{b}z^{N}+\textrm{other background fields}+\textrm{fermions}.\label{3.1} 
\end{eqnarray}
 We introduced the 5d variable \( z^{M}=(x^{\mu },y) \). We are also considering
a fermionic non-critical string. The metric \( G_{MN} \) has the form obtained
by the obvious change of variables in (\ref{1.1}):
\begin{equation}
\label{3.2}
ds^{2}=\rho (y)(dy^{2}+d\vec{x}\, ^{2})\; .
\end{equation}
 As usual, the background fields are determined from the \( \beta  \)-function
equation \cite{11}, expressing the \( g^{ab} \)-independence of the theory. 

It is important to have the NSR fermions for the gauge fields\,--\,strings duality.
It is also necessary to perform the non-chiral GSO-projection in the functional
integral. The whole matter has been treated in the paper \cite{12}. Here we
will just remind the reader the underlying logic with a few additional comments.

There are two types of objects we would like to consider. First, there are the
closed strings amplitudes. We introduce the closed string vertex operators,
e.g.
\[
V_{\mu \nu }=\int d^{2}\xi \left( \vphantom {\sum }\psi _{\vec{p}}(y(\xi ))\partial _{a}x^{\mu }\partial _{a}x^{\nu }\right) e^{i\vec{p}\, \vec{x}(\xi )}\]
 and claim that for the proper choice of the metric \( \rho (y) \) the correlators
of \( V \)'s on the string side are equal to the correlators of the gauge invariant
operators on the Yang-Mills side. In other words, we conjecture an isomorphism
\begin{equation}
\label{3.4}
\begin{array}{c}
\textrm{Closed string}\\
\textrm{states}
\end{array}\Longleftrightarrow \begin{array}{c}
\textrm{Gauge invariant}\\
\textrm{operators}.
\end{array}
\end{equation}
 The intuitive reason for this conjecture is that gauge invariant operators
like \( \textrm{Tr}F_{\mu \nu }^{2} \) can be associated with tiny closed flux
lines, represented by small Wilson loops. Indeed, we create a flux line by acting
on the vacuum with the operator 
\[
\textrm{Tr}P\exp \oint\limits _{C}A\, dx\mathop {\approx }_{C\to 0}I+C_{\mu \nu \lambda \sigma }\textrm{Tr}(F_{\mu \nu }F_{\lambda \sigma })+\ldots \]
 The propagation of these loops forms the closed world surfaces with punctures
represented by the vertex operators. This means that if we find the right action
(\ref{3.1}), the closed string \( S \)-matrix will be equal to the correlation
functions of the local gauge invariant operators.

But how to discover the right action? This question does not have a complete
answer at present. There are two approaches the problem. One approach \cite{2},\,\cite{3},
applicable to the supersymmetric Yang-Mills theories, is to start with the 3-branes
of the Type IIB superstring. Then, on the one hand, it is known that the low
energy excitation of the stack of \( N \) three-branes are described by the
supersymmetric Yang-Mills theory with the \( SU(N) \) gauge group. On the other
hand, 3-branes can be presumably replaced by the supergravity background the
generate, and thus one expects and checks the isomorphism (\ref{3.4}). 

The main problem with this approach is that it is far from being clear that
non-supersymmetric theories can be described in the D-brane language. Also,
even in the supersymmetric case, so far there is no real \emph{derivation} of
the isomorphism (\ref{3.4}) from the first principles, although numerous checks
have been performed.

Another approach \cite{1} is to try to formulate the conditions under which
the \( \sigma  \)-model satisfies the Schwinger-Dyson equations of the Yang-Mills
theory. The basic object in this approach is the Wilson loop, defined by
\[
W[\vec{x}(s)]=\int\limits _{\left\{ \begin{array}{c}
\vec{x}|_{\partial D}=\vec{x}(s)\\
y|_{\partial D}=?
\end{array}\right. }\mathcal{D}y(\xi )\mathcal{D}\vec{x}(\xi )e^{-S[\vec{x}(\xi ),y(\xi )]}\; .\]
 Here we take the world sheet with a disc topology (appropriate for large \( N \)).
At the boundary of the disc we fixed \( \vec{x} \) by the condition \( \vec{x}|_{\partial D}=\vec{x}(s) \).
Now, we have to determine \( y|_{\partial D} \) and the background fields so
that the loop equations are satisfied. The basic idea of \cite{1} was to notice
that the metric \( \rho (y) \) must have horizon and/or infinity points, so
that
\[
\rho (y_{H})=0\quad \textrm{and}\quad \rho (y_{I})=\infty \; .\]
 These are the only positions where we can place the contour. Indeed, if we
choose some other \( y|_{\partial D} \), we will encounter the lack of the
zigzag symmetry coming from the fact that the first term in (\pageref{3.1})
is (due to the presence of \( \sqrt{g} \)) invariant only under diffeomorphisms,
but not under the zigzag transformations. Hence it must be either zero or infinite
at the boundary. A more precise condition was given in \cite{12}. It was stated
there that in order to describe gauge theory we must find the metric \( \rho (y) \),
the other background fields, and the boundary conditions \( y|_{\partial D} \)
such that \emph{the vector vertex operators, defined at the boundary, form a
closed algebra}. Physically this means that open strings in this theory correspond
to vector gluons. Also, the vector vertex operators are selected by the zigzag
symmetry, since they are the only ones defined without the world sheet metric.

When the D-brane description is applicable, the above condition is satisfied.
But it is important that the condition be formulated without any reference to
D-branes. 

In the case of the \( \mathcal{N}=4 \) SYM the D-branes and conformal symmetry
considerations lead to the formula \cite{3}
\begin{equation}
\label{3.7}
\rho (y)=\sqrt{{g_{YM}^{2}N}}\cdot \frac{1}{y^{2}}\; .
\end{equation}
 It has been shown in \cite{4}, \cite{5} for the local operators and in \cite{13},\,\cite{14}
for the loops that the natural location for the gauge theory quantities is the
infinity in this AdS space, that is \( y\to 0 \). 

In this paper we shall adopt this prescription, although there is no real proof
for it. In a sense, our check of the loop equations in this paper can be considered
as such (incomplete) proof.

One should keep in mind, however, that there is another zigzag-symmetric location,
\( y=\infty  \). This point has no geometrical significance in the Euclidean
signature (the Lobachevsky space) but is meaningful in the Lorentz signature
(that is in the AdS space). In the latter case this point represents a horizon
behind which the D-branes are hidden. This horizon is physically important because
the absorption by the D-branes can be accounted for as the disappearance of
the closed strings behind the horizon \cite{2}. It is therefore natural that
the Wilson loop \( C \) on the D-brane can be represented as a Wilson loop
at the horizon \cite{1}. In this case, however, the loop generates the \( B_{\mu \nu } \)-field
with \( B_{\mu \nu }=B_{\mu \nu }(x,\{C\}) \) \cite{1}. At present we do not
know how to determine this field. We hope that the two types of boundary conditions
yield the same result and will be studying the case \( y=0 \).

The string representation involves various background fields. For example, since
we need NSR fermions to exclude the boundary tachion (violating the zigzag symmetry),
we have RR fields in our closed string spectrum, and they form a condensate
in order to stabilize the AdS space. The concrete types and forms of background
fields depend on the type of the gauge theory we are working with. The problem
of finding the precise background in each particular case is not yet completely
solved. Of course to check the complete loop equations we have to solve the
above problem first. However, in the conformal cases it is possible to retreat
to the quasiclassical domain by taking \( g_{YM}^{2}N\gg 1 \). In this case
the string action takes the form:
\begin{equation}
\label{3.8}
S[\vec{x}(\xi ),y(\xi )]=\frac{1}{2}\sqrt{g_{YM}^{2}N}\int \frac{d^{2}\xi }{y^{2}(\xi )}\left( (\partial _{a}\vec{x})^{2}+(\partial _{a}y)^{2}\right) +O(1)\; ,
\end{equation}
 where the terms \( O(1) \) contain NSR fermions and RR fields. We arrive at
the conclusion that in the WKB limit, the Wilson loop is given by \cite{13},\emph{\,}\cite{14}
\begin{equation}
\label{3.75}
W[C]\varpropto e^{-\sqrt{g_{YM}^{2}N}\cdot A_{\min }[C]}\; .
\end{equation}
 Here
\begin{equation}
\label{3.9}
A_{\min }[C]=\min \frac{1}{2}\int \frac{d^{2}\xi }{y^{2}}\left( (\partial _{a}\vec{x})^{2}+(\partial _{a}y)^{2}\right) 
\end{equation}
 is the minimal area of a surface bounded by the loop \( C \). In the following
sections we will analyze the action of the loop operator on this functional.

\section{Minimal area in the Lobachevsky space}

Our contour is located at infinity (the absolute) of the Lobachevsky space.
There are some general features of such minimal areas which we discuss in this
section. The equations of motion for the action (\ref{3.9}) have the form:
\begin{equation}
\label{4.1}
\left\{ \begin{array}{l}
\dis \partial _{a}\left( \frac{1}{y^{2}}\partial _{a}\vec{x}\right) =0\; ,\\
\dis \partial ^{2}y=\frac{1}{y}\left( (\partial _{a}y)^{2}-\left( \partial _{a}\vec{x}\right) ^{2}\right) \; .
\end{array}\right. 
\end{equation}
 We also have to impose the Virasoro constraints
\begin{equation}
\label{4.2}
\left\{ \begin{array}{l}
\dis \left( \partial _{1}\vec{x}\right) ^{2}+\left( \partial _{1}y\right) ^{2}=\left( \partial _{2}\vec{x}\right) ^{2}+\left( \partial _{2}y\right) ^{2}\; ,\\
\dis \partial _{1}\vec{x}\, \partial _{2}\vec{x}+\partial _{1}y\, \partial _{2}y=0\; .
\end{array}\right. 
\end{equation}
 We are looking for the solutions satisfying the conditions
\[
\left\{ \begin{array}{l}
\vec{x}(\sigma ,0)=\vec{c}(\sigma )\; ,\\
y(\sigma ,0)=0\; .
\end{array}\right. \]
 (We renamed the variables: \( \xi ^{1}=\sigma  \), \( \xi ^{2}=\tau  \).)
It is easy to see that the expansion in \( \tau  \) has the form:
\begin{equation}
\label{4.4}
\left\{ \begin{array}{l}
\dis \vec{x}=\vec{c}(\sigma )+\frac{1}{2}\vec{f}(\sigma )\tau ^{2}+\frac{1}{3}\vec{g}(\sigma )\tau ^{3}+\ldots \\
\dis y=a(\sigma )\tau +\frac{1}{3}b(\sigma )\tau ^{3}+\ldots 
\end{array}\right. 
\end{equation}
 After substituting this expansion into (\ref{4.1}) and (\ref{4.2}) we obtain
\begin{equation}
\label{4.5}
\left\{ \begin{array}{l}
\dis a^{2}(\sigma )=\left( \frac{d\vec{c}}{d\sigma }\right) ^{2}\; ,\\
\dis \vec{f}(\sigma )=\left( \frac{d\vec{c}}{d\sigma }\right) ^{2}\frac{d}{d\sigma }\left( \frac{\partial _{\sigma }\vec{c}}{\left( \partial _{\sigma }\vec{c}\, \right) ^{2}}\right) \; .
\end{array}\right. 
\end{equation}
 This guarantees that the leading term in the energy-momentum tensor (\ref{4.2})
vanishes; we have
\[
\theta _{\bot \parallel }=\frac{1}{y^{2}}\left( \partial _{\tau }\vec{x}\, \partial _{\sigma }\vec{x}+\partial _{\tau }y\, \partial _{\sigma }y\right) =\frac{1}{a^{2}\tau }\left[ \vec{f}{\vec{c}\, }'+aa'\right] +\frac{1}{a^{2}}\left( \vec{g}{\vec{c}\, }'\right) +\ldots \]
 and due to (\ref{4.5}) the first term vanishes, while the second one gives
\[
\theta _{\bot \parallel }=\frac{1}{a^{2}}\left( \vec{g}{\vec{c}\, }'\right) \; .\]
 Analogously:
\begin{eqnarray}
 & \theta _{\bot \bot } & =\dis \frac{1}{2y^{2}}\left[ \left( \partial _{\tau }\vec{x}\right) ^{2}+\left( \partial _{\tau }y\right) ^{2}-\left( \partial _{\sigma }\vec{x}\right) ^{2}-\left( \partial _{\sigma }y\right) ^{2}\right] \nonumber \\
 &  & =\dis \frac{1}{a^{2}}\left[ {\vec{f}\, }^{2}+2ab-{\vec{c}\, }'\vec{f}\, '-{a'}^{2}\right] \; .\label{4.8} 
\end{eqnarray}
 The action (\ref{3.9}) calculated on the classical solution (\ref{4.4}) has
the following structure (encountered previously in \cite{13} in a special case):
\begin{equation}
\label{4.9}
A_{\min }[C]=\frac{L[C]}{\epsilon }+\mathcal{A}[\vec{c}(\sigma )]\; ,
\end{equation}
 where we introduced the cut-off \( y_{\min }=a(\sigma )\tau _{\min }=\epsilon  \).
In this formula \( L[C] \) is the length of the contour \( C \) and \( \mathcal{A}[\vec{c}(\sigma )] \)
is a finite functional.

Our main interest is to derive variational equations for \( \mathcal{A}[C] \).
One might think that the usual Hamilton-Jacobi equations following from the
conditions \( \theta _{\bot \Vert }=\theta _{\bot \bot }=0 \) will give us
a closed equation for \( \mathcal{A} \). Unfortunately, life is not so simple.
The above would be the case is we were able to find the coefficients \( \vec{g} \)
and \( b \) in terms of \( \vec{c} \). Indeed, it is easy to see that
\[
\frac{\delta \mathcal{A}}{\delta \vec{c}(\sigma )}=\frac{\vec{g}(\sigma )}{a^{2}(\sigma )}\]
 and if \( b \) were known, the equation for \( \mathcal{A} \) would follow
from (\ref{4.8}).

There is an unpleasant surprise, however. Further iterations of (\ref{4.1})
reveal that the functions \( \vec{g}(\sigma ) \) and \( b(\sigma ) \) in (\ref{4.4})
are not determined by the small \( \tau  \)-expansion and can be kept arbitrary.
They are fixed by the global condition for the absence of singularities at finite
\( \tau  \), and thus it is hard to find them explicitly. Because of this difficulty
we will choose an alternative way of finding \( \mathcal{A}[C] \). Namely,
we will consider a special type of contours---wavy lines, and will develop a
method of successive approximations for \( \mathcal{A}[C] \). We will also
see another derivation of (\ref{4.9}).

\section{The theory of wavy lines}

It was already suggested in \cite{1} that it is instructive to consider wavy
lines instead of general contours, namely to look at a curve
\[
x^{1}(s)=s,\quad x^{i}(s)=\phi ^{i}(s),\quad i=2,\ldots D\; ,\]
 and to assume that \( \phi ^{i}(s) \) are small. Below we shall find the expansion
of \( \mathcal{A}[\vec{\phi }(s)] \) up to the fourth order and then explore
the action of the loop Laplacian. To perform this expansion it is convenient
to consider the standard Hamilton-Jacobi equation for the minimal surface which
has the well-known general form:
\[
G^{MN}(z)\frac{\delta A}{\delta z^{M}(s)}\frac{\delta A}{\delta z^{N}(s)}=G_{MN}\frac{dz^{M}}{ds}\frac{dz^{N}}{ds}\; .\]
 For the Poincar\'e metric this equation becomes
\begin{equation}
\label{5.3}
\left( \frac{\delta A}{\delta y(s)}\right) ^{2}+\left( \frac{\delta A}{\delta \vec{x}(s)}\right) ^{2}=\frac{1}{y^{4}(s)}\left\{ \left( \frac{dy}{ds}\right) ^{2}+\left( \frac{d\vec{x}}{ds}\right) ^{2}\right\} \; .
\end{equation}
 We want to explore the limit \( y(s)=y\to 0. \) We can do this by solving
(\ref{5.3}) with respect to \( \frac{\delta A}{\delta y} \) and by noticing
that
\begin{equation}
\label{5.4}
\frac{\partial A}{\partial y}=\int ds\frac{\delta A}{\delta y(s)}\Bigl |_{y(s)=y}\; .
\end{equation}
 We have from (\ref{5.3}) and (\ref{5.4})
\begin{equation}
\label{5.5}
\frac{\partial A}{\partial y}=-\frac{1}{y^{2}}\int ds\sqrt{{\left( \frac{d\vec{x}}{ds}\right) ^{2}-y^{4}\left( \frac{\delta A}{\delta \vec{x}(s)}\right) ^{2}}}\; .
\end{equation}
 We see directly from (\ref{5.5}) that \( A(y) \) behaves like
\[
A(y)\mathop {\approx }_{y\to 0}\frac{L[C]}{y}+O(1)\; .\]
 To get further information we have to look at the following expansion of \( A \):
\[
A=\sum _{n}\frac{1}{n!}\int ds_{1}\ldots ds_{n}\Gamma _{i_{1}\ldots i_{n}}(s_{1},\ldots s_{n}|y)\bigl (\phi _{i_{1}}(s_{1})\ldots \phi _{i_{n}}(s_{n})\bigl )\; .\]
 Expansion of (\ref{5.5}) together with the reparametrization invariance equation
\begin{equation}
\label{5.75}
\frac{dx_{\mu }}{ds}\frac{\delta A}{\delta x_{\mu }(s)}=\frac{\delta A}{\delta x_{1}(s)}\Bigl |_{x_{1}=s}+\frac{d\vec{\phi }}{ds}\frac{\delta A}{\delta \vec{\phi }}=0
\end{equation}
 will give us equations for the \( \Gamma  \)'s. Let us expand (\ref{5.5}):
\begin{eqnarray}
 & \dis \frac{\partial A}{\partial y} & =-\frac{L_{0}}{y^{2}}+\frac{1}{2}\int \left( y^{2}\vec{\pi }^{2}-\frac{1}{y^{2}}{\vec{\phi }\, '}^{2}\right) ds\nonumber \\
 &  & \qquad \qquad +\frac{1}{8}\int \left( \frac{1}{y^{2}}\left( y^{4}\vec{\pi }^{2}-{\vec{\phi }\, '}^{2}\right) ^{2}+4y^{2}\left( \vec{\phi }\, '\vec{\pi }\right) ^{2}\right) ds+\ldots \label{5.8} 
\end{eqnarray}
 where \( \vec{\pi }=\delta A/\delta \vec{\phi } \). If we perform the Fourier
transform in \( s \), we obtain the following equations for the coefficient
functions:
\begin{equation}
\label{5.9}
\left\{ \begin{array}{l}
\dis \frac{d\Gamma _{2}}{dy}=y^{2}\Gamma ^{2}_{2}-\frac{p^{2}}{y^{2}}\; ,\\
\dis \frac{d\Gamma _{4}}{dy}=y^{2}\left( \sum _{1}^{4}\Gamma _{2}(p_{i})\right) \Gamma _{4}(p_{1},\ldots p_{4})-B_{4}(p_{1},\ldots p_{4})\; .
\end{array}\right. 
\end{equation}
 We assume here that
\begin{eqnarray}
 & A & =\frac{L_{0}}{y}+\frac{1}{2}\int \Gamma _{2}(p)\left( \vec{\phi }_{p}\vec{\phi }_{-p}\right) \, dp-\label{5.10} \\
 &  & \qquad -\frac{1}{8}\int \Gamma _{4}(p_{1},\ldots p_{4})\left( \vec{\phi }_{p_{1}}\vec{\phi }_{p_{2}}\right) \left( \vec{\phi }_{p_{3}}\vec{\phi }_{p_{4}}\right) \delta \left( \sum p_{i}\right) dp_{1}\ldots dp_{4}+\ldots \nonumber 
\end{eqnarray}
 The only slightly complicated structure is the polynomial \( B_{4} \) which
is found from (\ref{5.8}):
\begin{eqnarray}
(2\pi )B_{4}(p_{1},\ldots p_{4}) & = & C_{4}(p_{1},\ldots p_{4})+D(p_{1},p_{2})+D(p_{3},p_{4})-\nonumber \\
 &  & \qquad -D(p_{1},p_{3})-D(p_{1},p_{4})-D(p_{2},p_{3})-D(p_{2},p_{3})\; .\label{5.11} 
\end{eqnarray}
 In this formula 
\begin{eqnarray}
C_{4} & = & \left( p_{1}p_{2}p_{3}p_{4}\right) \frac{1}{y^{2}}+y^{6}\left( \omega _{1}\omega _{2}\omega _{3}\omega _{4}\right) \; ,\nonumber \\
D(p_{1},p_{2}) & = & (p_{1}p_{2}\omega _{3}\omega _{4})y^{2}\; ,\nonumber \label{5.12} 
\end{eqnarray}
 where we introduced the notation \( \omega _{i}=\Gamma _{2}(p_{i}) \).

The equations (\ref{5.9}) are easy to solve. From the first one we get
\begin{equation}
\label{5.13}
\Gamma _{2}(p,y)\equiv \omega (p,y)=\frac{p^{2}}{y(1+|p|y)}
\end{equation}
 (this solution is the only one which is positive and regular for \( y\to \infty  \).)
From the second one,
\begin{eqnarray}
\Gamma _{4} & = & \int _{y}^{\infty }dy\, e^{-\int _{0}^{y}dy_{1}y_{1}^{2}\left( \sum \omega _{i}\right) }B_{4}(p_{1},\ldots p_{4}|y)=\nonumber \\
 & = & \int _{y}^{\infty }dy\, \prod _{i}(1+|p_{i}|y)e^{-\sum |p_{i}|y}B_{4}(p_{1},\ldots p_{4}|y).\label{5.14} 
\end{eqnarray}
 We see that \( \Gamma _{4} \) also has the structure of (\ref{5.11}):
\begin{equation}
\label{5.15}
(2\pi )\Gamma _{4}=F(p_{1},\ldots p_{4})+\Phi _{12}+\Phi _{34}-\Phi _{13}-\Phi _{14}-\Phi _{23}-\Phi _{24}.
\end{equation}
 Taking the integral (\ref{5.14}) and separating the finite part for \( y\to 0 \)
gives
\begin{eqnarray}
F & = & \left( 2\frac{\epsilon _{1}\epsilon _{2}\epsilon _{3}\epsilon _{4}+1}{\Delta ^{3}}+\frac{\epsilon _{1}\epsilon _{2}\epsilon _{3}\epsilon _{4}}{\Delta ^{2}}\left( \sum \frac{1}{|p_{i}|}\right) +\frac{\sum _{i<j}|p_{i}|\cdot |p_{j}|}{\Delta \, p_{1}p_{2}p_{3}p_{4}}-\frac{\Delta }{p_{1}p_{2}p_{3}p_{4}}\right) p_{1}^{2}p_{2}^{2}p_{3}^{2}p_{4}^{2},\nonumber \\
\Phi _{12}\! \! \! \!  & = & \left( \frac{2\epsilon _{1}\epsilon _{2}}{\Delta ^{3}}+\frac{\epsilon _{1}\epsilon _{2}}{\Delta ^{2}}\left( \frac{1}{|p_{1}|}+\frac{1}{|p_{2}|}\right) +\frac{1}{\Delta \, p_{1}p_{2}}\right) p_{1}^{2}p_{2}^{2}p_{3}^{2}p_{4}^{2}.\label{5.17} 
\end{eqnarray}
 (\( \Delta =\sum |p_{i}| \); \( \epsilon _{i}=\textrm{sgn }p_{i} \)). This
completes the calculation of \( \Gamma _{4}(p_{1},\ldots p_{4}) \). 

Let us notice that an alternative way to obtain these formulas is to use the
Monge gauge in the expression for the minimal area: \( x_{1}=\sigma ,\, y=\tau ,\, \phi _{i}=\phi _{i}(\sigma ,\tau ), \)
\begin{equation}
\label{5.18}
A=\int \frac{d\tau }{\tau ^{2}}\sqrt{{1+\vec{\phi }_{\tau }^{\, 2}+\vec{\phi }_{\sigma }^{\, 2}+\vec{\phi }_{\tau }^{\, 2}\vec{\phi }_{\sigma }^{\, 2}-\left( \vec{\phi }_{\tau }\vec{\phi }_{\sigma }\right) ^{2}}}.
\end{equation}
 The linear equation for \( \vec{\phi } \),
\[
\partial _{\tau }\left( \frac{1}{\tau ^{2}}\partial _{\tau }\vec{\phi }\right) +\frac{1}{\tau ^{2}}\partial _{\sigma }^{2}\vec{\phi }=0\]
 has the solution
\[
\vec{\phi }_{\textrm{cl}}(p,\tau )=(|p|\tau )^{3/2}K_{3/2}(|p|\tau )\vec{\phi }(p)=(1+|p|\tau )e^{-|p|\tau }\vec{\phi }(p).\]
Expanding (\ref{5.18}), we arrive once again at the expression (\ref{5.14}).
However, the Hamilton-Jacobi method has some general advantages.

\section{The second derivative of the minimal area}

After the divergent part of (\ref{4.9}) is absorbed into the mass renormalization
of the test particle, we are left with the functional 
\begin{equation}
\label{7.05}
W[C]=e^{-\sqrt{{g^{2}_{YM}N}}\mathcal{A}[C]}\: .
\end{equation}
 Its second variational derivative has the form
\[
\frac{\delta ^{2}W}{\delta x_{\mu }(s)\delta x_{\mu }(s')}=\left( g^{2}_{YM}N\frac{\delta \mathcal{A}}{\delta x_{\mu }(s)}\frac{\delta \mathcal{A}}{\delta x_{\mu }(s')}-\sqrt{{g^{2}_{YM}N}}\frac{\delta ^{2}\mathcal{A}}{\delta x_{\mu }(s)\delta x_{\mu }(s')}\right) W\; .\]
 The first term in brackets has no singularity for \( s\to s' \). This fact
was considered in \cite{15} as a check that the loop equation is satisfied.
Notice however that in this order an arbitrary functional \( \mathcal {A} \)
will pass this check. That is why it is necessary to consider the second term,
which we proceed to calculate. 

The full second variational derivative of \( \mathcal {A} \) consists of the
longitudinal and transverse parts:
\begin{equation}
\label{7.1}
\frac{\delta ^{2}\mathcal{A}}{\delta x_{\mu }(s)\delta x_{\mu }(s')}=\frac{\delta ^{2}\mathcal{A}}{\delta x_{1}(s)\delta x_{1}(s')}+\frac{\delta ^{2}\mathcal{A}}{\delta \vec{\phi }(s)\delta \vec{\phi }(s')}\; .
\end{equation}
The transverse part (in momentum representation) can be read off directly from
(\ref{5.10}). For the term quadratic in \( \phi  \) we have 
\[
\lim _{p\to \infty }\frac{\delta ^{2}\mathcal{A}^{(2)}}{\delta \vec{\phi }\left( \frac{q}{2}+p\right) \delta \vec{\phi }\left( \frac{q}{2}-p\right) }=(D-1)\delta (q)\Gamma _{2}(p)=(1-D)\delta (q)|p|^{3}\; ,\]
where \( \Gamma _{2}(p)=-|p|^{3} \) is the finite part of \( \Gamma _{2}(p,y) \). 

Let us now consider the quartic term. The special structure in (\ref{5.15})
leads to the formula
\begin{eqnarray}
 &  & (2\pi )\frac{\delta ^{2}\mathcal{A}^{(4)}}{\delta \vec{\phi }(k)\delta \vec{\phi }(k')}=-\frac{1}{2}\int H(k,k',p_{1},p_{2})\left( \vec{\phi }_{p_{1}}\vec{\phi }_{p_{2}}\right) \, \delta (k+k'+p_{1}+p_{2})\, dp_{1}dp_{2}\; ,\nonumber \\
 &  & H=(D+1)F(k,k',p_{1},p_{2})+(D-3)\bigl [\Phi (k,k')+\Phi (p_{1},p_{2})\bigr ]\nonumber \\
 &  & \qquad \qquad \qquad \qquad \qquad -(D-1)\bigr [\Phi (k,p_{1})+\Phi (k,p_{2})+\Phi (k',p_{1})+\Phi (k',p_{2})\bigl ]\; .\label{7.5} 
\end{eqnarray}
 Let us first analyze the case \( k=-k' \), which corresponds to taking \( q=0 \)
in (\ref{2.7}). In this case the third term in the expression for \( H \)
in (\ref{7.5}) drops out, because \( \Phi (k,p) \) is antisymmetric with respect
to \( k\to -k \). The remaining terms have to be expanded for \( k\to \infty . \)
Because of homogeneity, it makes sense to express them as functions of \( x=k/p \),
where \( p_{1}=-p_{2}=p \), which amounts to taking \( p=1 \) in (\ref{5.17}).
We have: 
\begin{eqnarray}
F(k,-k,p,-p) & = & |p|^{5}\Bigl \{\frac{x^{4}}{2(1+x)^{3}}+\frac{x^{3}}{2(1+x)}+\frac{x^{2}(1+4x+x^{2})}{2(1+x)}-2x^{2}(1+x)\Bigr \}\nonumber \\
 & = & |p|^{5}\Bigl \{-\frac{3}{2}x^{3}-x+O\left( \frac{1}{x}\right) \Bigl \}\; ,\nonumber \\
\Phi (k,-k)+\Phi (p,-p) & = & |p|^{5}\Bigl \{-\frac{x^{2}(1+x^{2})}{2(1+x)}-\frac{x^{3}}{2(1+x)}-\frac{x^{4}}{2(1+x)^{3}}\Bigr \}\nonumber \\
 & = & |p|^{5}\Bigr \{-\frac{1}{2}x^{3}-x+2+O\left( \frac{1}{x}\right) \Bigl \}\; .\nonumber \label{7.6} 
\end{eqnarray}
  We substitute this into (\ref{7.5}) and obtain the desired asymptotic expansion
in \( k\to \infty  \):
\begin{eqnarray}
\frac{\delta \mathcal{A}^{(4)}}{\delta \vec{\phi }(k)\delta \vec{\phi }(-k)} & = & \frac{1}{2\pi }\int \Bigl \{Dp^{2}|k|^{3}+(D-1)p^{4}|k|\nonumber \\
 &  & \qquad \qquad +(3-D)|p|^{5}\Bigl \}\left( \vec{\phi }_{p}\vec{\phi }_{-p}\right) \, dp+O\left( \frac{1}{k}\right) \; .\label{7.9} 
\end{eqnarray}
In general, when we do not assume that \( q=0 \) in (\ref{2.7}), we have to
take all the terms in (\ref{7.5}) into account. The corresponding formula for
the asymptotic expansion (of which (\ref{7.9}) is a partial case) can be obtained
by means of straightforward but lengthy calculations and has the form:
\begin{eqnarray}
 &  & \frac{\delta ^{2}\mathcal{A}^{(4)}}{\delta \vec{\phi }\left( \frac{q}{2}+k\right) \delta \vec{\phi }\left( \frac{q}{2}-k\right) }=\frac{1}{2\pi }\int \Bigl \{-Dp_{1}p_{2}|k|^{3}\nonumber \\
 &  & \qquad \qquad +\left( \frac{D-4}{2}p_{1}p_{2}^{3}+\frac{3D-6}{2}p_{1}^{2}p_{2}^{2}\right) |k|+\left( (4-D)p_{1}^{2}|p_{2}|^{3}+p_{1}p_{2}|p_{2}|^{3}\right) \Bigl \}\nonumber \\
 &  & \qquad \qquad \qquad \qquad \qquad \times \left( \vec{\phi }_{p_{1}}\vec{\phi }_{p_{2}}\right) \, \delta (p_{1}+p_{2}+q)\, dp_{1}dp_{2}+O\left( \frac{1}{k}\right) \; .\label{7.10} 
\end{eqnarray}

To complete the calculation of (\ref{7.1}), we have to compute the longitudinal
part (the first term in (\ref{7.1})). This part is not immediately visible
from (\ref{5.10}) and has to be recovered by the use of the reparametrization
invariance. The required identity can be easily derived from (\ref{5.75}) and
has the form:
\[
\frac{\delta ^{2}\mathcal{A}}{\delta x_{1}(s)\delta x_{1}(s')}=\dot{\phi }_{i}(s)\dot{\phi }_{k}(s')\frac{\delta ^{2}\mathcal{A}}{\delta \phi _{i}(s)\delta \phi _{k}(s')}-\delta (s-s')\dvp (s)\frac{d}{ds}\left( \frac{\delta A}{\delta \vec{\phi }(s)}\right) \; .\]
 In the momentum representation:
\begin{eqnarray}
(2\pi )\frac{\delta ^{2}\mathcal{A}}{\delta x_{1}\left( k\right) \delta x_{1}\left( k'\right) }=-\int dp\, dp'\, p\, p'\phi _{i}(p)\phi _{k}(p')\frac{\delta ^{2}\mathcal{A}}{\delta \phi _{i}(p+k)\delta \phi _{k}(p'+k')} &  & \nonumber \\
-\int dp\, p(p+k+k')\vec{\phi }(p)\frac{\delta \mathcal{A}}{\delta \vec{\phi }(p+k+k')}\; . &  & \label{7.15} 
\end{eqnarray}
 In the approximation we are working with, only \( \mathcal{A}^{(2)} \) contributes
to the RHS of (\ref{7.15}). The result is:
\begin{eqnarray}
\frac{\delta ^{2}\mathcal{A}}{\delta x_{1}\left( \frac{q}{2}+k\right) \delta x_{1}\left( \frac{q}{2}-k\right) } & = & \frac{1}{2\pi }\int\limits \Bigl \{\Bigl |\frac{p_{1}-p_{2}}{2}+k\Bigr |^{3}-|p_{2}|^{3}\Bigl \}\nonumber \\
 &  & \qquad \qquad \times p_{1}p_{2}\left( \vec{\phi }_{p_{1}}\vec{\phi }_{p_{2}}\right) \delta (p_{1}+p_{2}+q)\, dp_{1}dp_{2}\nonumber \\
 & \mathop {=}_{k\to \infty } & \frac{1}{2\pi }\int \Bigl \{|k|^{3}+\frac{3}{4}(p_{1}-p_{2})^{2}|k|-p_{1}p_{2}|p_{2}|^{3}\Bigl \}\times \ldots \label{7.16} 
\end{eqnarray}
 The final expression for the asymptotics of the second variational derivative
is thus:
\begin{eqnarray}
 &  & \frac{\delta ^{2}\mathcal{A}^{(4)}}{\delta x_{\mu }\left( \frac{q}{2}+k\right) \delta x_{\mu }\left( \frac{q}{2}-k\right) }=(1-D)\delta (q)|k|^{3}+\frac{1}{2\pi }\int \Bigl \{(1-D)p_{1}p_{2}|k|^{3}\nonumber \\
 &  & \qquad \qquad +\left( \frac{D-1}{2}p_{1}p_{2}^{3}+\frac{3D-9}{2}p_{1}^{2}p_{2}^{2}\right) |k|+(4-D)p_{1}^{2}|p_{2}|^{3}\Bigl \}\nonumber \\
 &  & \qquad \qquad \qquad \qquad \times \left( \vec{\phi }_{p_{1}}\vec{\phi }_{p_{2}}\right) \, \delta (p_{1}+p_{2}+q)\, dp_{1}dp_{2}+O\left( \frac{1}{k}\right) +O(\phi ^{4})\; .\label{7.17} 
\end{eqnarray}
 In the next section we will analyze this result.

\section{Interpretation and discussion of the result}

Our main result is contained in the formula (\ref{7.17}). It shows, first of
all, that the second variational derivative has the expected form (\ref{2.6}).
Namely, dangerous terms \( \propto \delta ''(s-s') \) (which would manifest
themselves in (\ref{7.17}) as terms \( \propto k^{2} \)) cancel for all \( D \).
The presence of those terms would imply that our functional is \emph{not} zigzag-invariant,
that is not presentable in the form (\ref{2.8}). 

The next result following from (\ref{7.17}) concerns the loop operator \( \widehat{{L}}_{q} \).
Using (\ref{2.7}) and picking up \( k^{0} \) terms in (\ref{7.17}), we get
\begin{equation}
\label{8.1}
\widehat{{L}}_{q}\mathcal{A}=\frac{4-D}{2\pi }\int p_{1}^{2}|p_{2}|^{3}\left( \vec{\phi }_{p_{1}}\vec{\phi }_{p_{2}}\right) \, \delta (p_{1}+p_{2}+q)\, dp_{1}dp_{2}\; .
\end{equation}
This shows that at \( D=4 \) the loop equation is satisfied (at least in our
approximation)! We will discuss the significance of this fact in the next section.

Now let us perform another test. Consider once again the OPE (\ref{2.4}) and
(\ref{2.6}) and let us try to determine its contribution to (\ref{7.17}).
If we assume that in conformal theory in the WKB limit the field strength keeps
its normal dimension 2 (this will be actually more of the conclusion than of
the assumption), we get
\begin{eqnarray}
F_{\mu \lambda }(x_{1})F_{\mu \sigma }(x_{2}) & \sim  & C_{1}\frac{\delta _{\lambda \sigma }}{|x_{1}-x_{2}|^{4}}+C_{2}\frac{(x_{1}-x_{2})_{\lambda }(x_{1}-x_{2})_{\sigma }}{|x_{1}-x_{2}|^{6}}\nonumber \\
 &  & +\, C_{3}\frac{(x_{1}-x_{2})_{(\lambda }(x_{1}-x_{2})_{\mu }F_{\mu \sigma )}(x)}{|x_{1}-x_{2}|^{4}}+C_{4}\frac{(x_{1}-x_{2})_{\mu }\nabla _{\mu }F_{\lambda \sigma }(x)}{\left| x_{1}-x_{2}\right| ^{2}}\; .\label{8.2} 
\end{eqnarray}
(\( x=\frac{x_{1}+x_{2}}{2} \)). The last two terms give no singular contribution
to (\ref{2.2}). The first two give after some calculations:
\[
\frac{\delta ^{2}W}{\delta x_{\mu }(s)\delta x_{\mu }(s')}\mathop {=}_{s\to s'}\frac{1}{|s-s'|^{4}}\frac{C_{1}+C_{2}}{\dot{x}^{2}}+\frac{1}{|s-s'|^{2}}\left( \frac{(C_{1}+C_{2})(\dot{x}\! \stackrel{{\, ...}}{x})}{12\dot{x}^{4}}+\frac{C_{1}\ddot{x}^{2}}{4\dot{x}^{4}}+\frac{C_{2}(\dot{x}\ddot{x})^{2}}{4\dot{x}^{4}}\right) \; .\]
The derivatives in the RHS are taken at the point \( \bar{s}=\frac{s+s'}{2}. \)
In the wavy line approximation we get
\begin{equation}
\label{8.35}
\frac{\delta ^{2}W}{\delta x_{\mu }(s)\delta x_{\mu }(s')}\mathop {=}_{s\to s'}\frac{1}{|s-s'|^{4}}(C_{1}+C_{2})(1-\dot{\phi }^{2})+\frac{1}{|s-s'|^{2}}\left( \frac{(C_{1}+C_{2})}{12}\dot{\phi }\stackrel{{\, ...}}{\phi }+\frac{C_{1}}{4}\ddot{\phi }^{2}\right) +O(\phi ^{4})\; .
\end{equation}
 We have to compare this behavior with our formula (\ref{7.17}). Picking up
the terms \( \propto |k|^{3} \) and \( |k| \) and using the Fourier transform
identities
\[
|k|^{3}\leftrightarrow \frac{1}{|s-s'|^{4}},\qquad |k|\leftrightarrow -\frac{1}{6|s-s'|^{2}}\]
 to go back to the \( s \)-representation, we have
\begin{equation}
\label{8.4}
\frac{\delta ^{2}\mathcal{A}}{\delta x_{\mu }(s)\delta x_{\mu }(s')}\propto \frac{1}{|s-s'|^{4}}(1-D)(1-\dot{\phi }^{2})+\frac{1}{|s-s'|^{2}}\left( \frac{1-D}{12}\dot{\phi }\stackrel{{\, ...}}{\phi }+\frac{3-D}{4}\ddot{\phi }^{2}\right) .
\end{equation}
 We see that this formula can be put in complete agreement with (\ref{8.35})
by taking \( C_{1}=D-3 \), \( C_{2}=2 \). The second term in (\ref{8.4})
may be modified if the theory contains a scalar operator of dimension 2. That
can change the constants \( C_{1} \) and \( C_{2} \). At the same time, the
structure of the first term cannot be modified by anything and provides a strong
check for the consistency of our approach. Notice that it also \emph{predicts}
that the dimension of \( F_{\mu \nu } \) is not renormalized.

\section{Conclusions and outlook}

The main efforts of this work were directed towards the development of new techniques
for checking the loop equations in string theory. We managed to apply the loop
Laplacian to the minimal area in the Lobachevsky space and to show for the first
time that the equations of motion of gauge theory are satisfied by string theory,
at least in the WKB approximation. This point perhaps requires some clarifications.
Namely, we looked at the Wilson loop (\ref{2.1}) in conformal versions of the
Yang-Mills theory. These versions unavoidably contain other fields. Hence we
expect that 
\[
\nabla _{\mu }F_{\mu \nu }=J_{\nu }\ne 0\]
(where \( J_{\nu } \) is the current generated by those fields). So, what is
the meaning of finding that \( \widehat{L}A_{\min }=0 \) at \( D=4 \)? 

There are several possible interpretations of this fact. Let us begin with the
unpleasant one (in which we do not believe). It may be that our result is just
a fluke, while if we proceed to higher orders in ``waviness'' of our contour,
the loop equations will not be satisfied. Further progress will be difficult
in this case.

Let us take another, optimistic view. Various gauge theories presumably have
a string representation with the background (\ref{1.1}). If the theory is conformal,
(\ref{1.1}) must describe the AdS space. What distinguishes various theories
is not the metric but other background fields and also the field content on
the world sheet. However, in the WKB limit \( g_{YM}^{2}N\to \infty  \) the
asymptotics of the Wilson loop is given by the formula (\ref{3.75}) and is
the same for all conformal theories. If this is the case, it means that the
current in the above formula is negligible in the WKB limit, and our result
\emph{actually checks the universal law} (\ref{3.75}). Notice also that we
are considering the standard Wilson loop and not its modified version suggested
in \cite{13},\,\cite{14}. Once again, with the above philosophy this modification
is irrelevant in the WKB limit. 

To verify the above assertions it is necessary to go beyond our wavy line approximation.
We believe that this can be done by a more general treatment of the Hamilton-Jacobi
equations. Conformal invariance of the loop Laplacian (which we discuss in the
Appendix) should play an important role in this analysis. Alternatively, one
can study the second variation of the functional (\ref{3.8}) by developing
the short distance expansion of the Green functions for equations (\ref{4.1}).
It is conceivable that by this method it will be possible to relate OPE on the
world sheet and OPE in gauge theory. 

This brings us to a more difficult problem of going beyond WKB approximation.
Again, the method of wavy lines may be useful here, but at the moment we do
not know how to evaluate the action of \( \widehat{L}_{q} \) on quantum corrections
to our formula. 

In the case of non-conformal theories the metric has the form
\[
ds^{2}=f(\log y)\left( \frac{dy^{2}+d\vec{x}\, ^{2}}{y^{2}}\right) \]
 The problem for our analysis is not so much the function \( f \) (it is easy
to generalize our considerations to this case) as the absence of the WKB domain.
It may be helpful to notice that instead of considering non-conformal case in
4d, one can, in the asymptotically free theories, shift to \( D=4+\epsilon  \),
when these theories become conformal. Perhaps, the classical part of \( \widehat{L}_{q}W\propto \epsilon  \)
will be canceled by the quantum fluctuations and will provide us with the ``Holy
Grail'' of this subject\,---\,the space-time \( \beta  \)-function.

\section*{Acknowledgements}

We are grateful to Volodya Kazakov for his participation in the early stages
of this project as well as for useful discussions. The work of A.P. was partially
supported by NSF grant PHY-98-02484.

\section*{Appendix: Conformal invariance of the loop equation}

The basic property of the functional \( \mathcal{A} \) (and hence of \( W \))
is its conformal invariance: 
\begin{equation}
\label{9.5}
\mathcal{A}[C]=\mathcal{A}[f(C)]\; 
\end{equation}
 (where \( f \) is a conformal transformation, say \( f_{\mu }(x)=x_{\mu }/x^{2} \)).
This is true for any \( D \). Although this invariance is to be expected, it
is not entirely obvious, since isometries of the Lobachevsky space (corresponding
to conformal transformations on the boundary) will change the cut-off \( \epsilon  \)
in (\ref{4.9}). To check the invariance, we extend \( f \) to the \( (D+1) \)-dimensional
Lobachevsky isometry:
\begin{equation}
\label{9.2}
(x_{\mu },y)\stackrel{{F}}{\rightarrow }\left( \frac{x_{\mu }}{x^{2}+y^{2}},\frac{y}{x^{2}+y^{2}}\right) \; .
\end{equation}
Consider the minimal surface \( M \) bounded by \( C \). Since \( F \) is
an isometry, \( F(M) \) is the corresponding surface for \( f(C) \). Moreover,
\begin{equation}
\label{9.3}
\textrm{Area}[M_{\epsilon }]=\textrm{Area}[F(M_{\epsilon })]
\end{equation}
(where \( M_{\epsilon } \) is the surface \( M \) cut off at the hight \( \epsilon  \)).
For small \( \epsilon  \), the LHS is equal to \( L[C]/\epsilon +\mathcal{A}[C] \).
According to (\ref{9.2}), the constant cut-off \( \epsilon  \) is transformed
by \( F \) to the variable cut-off \( \epsilon /x^{2} \). To calculate the
RHS of (\ref{9.3}), it is convenient to introduce an auxiliary constant cut-off
\( \epsilon ' \) on \( F(M) \), so that Area\( [F(M_{\epsilon })] \) becomes
equal to Area\( [F(M)_{\epsilon '}] \) plus the area of a narrow strip between
the variable and constant cut-offs. Now (\ref{9.3}) implies:
\[
\frac{L[C]}{\epsilon }+\mathcal{A}[C]\mathop {=}_{\epsilon ,\epsilon '\to 0}\frac{L[f(C)]}{\epsilon '}+\mathcal{A}[f(C)]+\oint _{C}\left| \frac{df(x(s))}{ds}\right| ds\int _{\epsilon /x(s)^{2}}^{\epsilon '}\frac{dy}{y^{2}}\; .\]
 Calculating the integral, we see that the singular terms cancel, and we get
(\ref{9.5}). 

It is now natural to ask if the loop Laplacian transforms is conformally invariant,
i.e. commutes with conformal transformations. This property can be written as
the equality
\begin{equation}
\label{9.6}
\widehat{{L}}(s)U_{f}[C]=\rho \left( \widehat{{L}}(s)U\right) [f(C)]
\end{equation}
 valid for any (reparametrization invariant) functional \( U[C] \) (where \( U_{f}[C]=U[f(C)] \)
is the functional \( U \) transformed by a conformal transformation; \( \rho  \)
is some factor). It turns out that (\ref{9.6}) is true if and only if \( D=4 \).
To prove this, consider the relation:
\[
\frac{\delta ^{2}U_{f}[x(s)]}{\delta x_{\mu }(s)\delta x_{\mu }(s')}=\partial _{\mu }f_{\lambda }(x(s))\partial _{\mu }f_{\sigma }(x(s'))\frac{\delta ^{2}U}{\delta f_{\lambda }(s)\delta f_{\sigma }(s')}+\partial ^{2}f_{\lambda }(x(s))\frac{\delta U}{\delta f_{\lambda }(s)}\delta (s-s')\; .\]
 For conformal transformations we have
\[
\partial _{\mu }f_{\lambda }\partial _{\mu }f_{\sigma }=\rho (f)\delta _{\lambda \sigma }\; .\]
 Now, we must collect terms proportional to \( \delta (s-s') \). This gives
\begin{eqnarray*}
 &  & \widehat{{L}}(s)U_{f}[C]\delta (s-s')=\rho (f)\left( \widehat{{L}}(s)U\right) [f(C)]\delta (s-s')\nonumber \\
 &  & \qquad \qquad +\, \partial ^{2}f_{\lambda }\frac{\delta U}{\delta f_{\lambda }}\delta (s-s')+\partial _{\mu }f_{\lambda }(x(s))\partial _{\mu }f_{\sigma }(x(s'))N_{[\lambda \sigma ]}\delta '(s-s')\; ,\label{6.3} 
\end{eqnarray*}
 where \( N_{[\lambda \sigma ]} \) is the coefficient in front of the \( \delta ' \)-function
in the second variational derivative:
\[
\frac{\delta ^{2}U}{\delta f_{\lambda }(s)\delta f_{\sigma }(s')}=N_{[\lambda \sigma ]}\Bigl (\frac{s+s'}{2}\Bigr )\delta '(s-s')+\ldots \]
 From the condition of the reparametrization invariance \( \frac{dx_{\lambda }}{ds}\frac{\delta U}{\delta x_{\lambda }(s)}=0 \)
we have the identity \cite{6}:
\[
\frac{\delta U}{\delta x_{\lambda }(s)}=N_{[\lambda \sigma ]}(s)\dot{x}_{\sigma }(s)\; .\]
 This gives the transformation law
\begin{eqnarray}
\widehat{{L}}(s)U_{f}[C] & = & \rho (f)\left( \widehat{{L}}(s)U\right) [f(C)]+\Omega _{[\lambda \sigma ]\mu }(f)\, N_{[\lambda \sigma ]}\, \dot{x}_{\mu }(s)\; ,\nonumber \\
\Omega _{[\lambda \sigma ]\mu } & = & \left( \partial ^{2}f_{[\lambda }\right) \left( \partial _{\mu }f_{\sigma ]}\right) -\left( \partial _{\alpha }\partial _{\mu }f_{[\lambda }\right) \left( \partial _{\alpha }f_{\sigma ]}\right) \; .\label{9.10} 
\end{eqnarray}
 Substituting \( f_{\mu }=x_{\mu }/x^{2} \), we find that \( \Omega (f)\varpropto (D-4) \),
and thus (\ref{9.6}) is true if and only if \( D=4 \). 

As a consequence of the above discussion, for \( D=4 \) the loop equation for
the Wilson loop (\ref{7.05}) is conformally invariant:
\begin{equation}
\label{9.1}
\widehat{L}(s)W[f(C)]=\rho \widehat{L}(s)W[C]\; .
\end{equation}
 It follows that in order to check the equation at a point \( x(s) \) of a
contour \( C \), we are allowed to first apply a conformal transformation,
say with the purpose of simplifying the behavior of the contour at the point
we are looking at. Although this observation does not play any significant role
when working with wavy lines, it might become important for general contours. 

For \( D\ne 4 \) the presence of nonzero additional term in the RHS of the
transformation law (\ref{9.10}) implies that the loop equation for (\ref{7.05})
\emph{cannot} be satisfied in this case.


\begin{thebibliography}{10}
\bibitem{1}A. Polyakov, Proceedings of ``Strings 97'', hep-th/9711002
\bibitem{2}I. Klebanov, Nucl. Phys. B496 (1997) 231, hep-th/9702076
\bibitem{3}J. Maldacena, Adv. Theor. Math. Phys. 2 (1998) 231, hep-th/9711200
\bibitem{4}S. Gubser, I. Klebanov, A. Polyakov, Phys. Lett. B428 (1998) 105, hep-th/9802109
\bibitem{5}E. Witten, Adv. Theor. Math. Phys. 2 (1998) 253, hep-th/9802150
\bibitem{6}A. Polyakov, Nucl. Phys. B164 (1980) 171
\bibitem{7}A. Polyakov, Phys. Lett. B82 (1979) 247
\bibitem{8}Yu. Makeenko, A. Migdal, Phys. Lett. B88 (1979) 135
\bibitem{9}Y. Nambu, Phys. Lett. B80 (1979) 372
\bibitem{10}J. Gervais, A. Neveu, Phys. Lett. B80 (1979) 255
\bibitem{11}C. Callan, E. Martinec, M. Perry, D. Friedlan, Nucl. Phys. B262 (1985) 593
\bibitem{12}A. Polyakov, J. Mod. Phys. A14 (1999) 645, hep-th/9809057
\bibitem{13}J. Maldacena, Phys. Rev. Lett. 80 (1998) 4859, hep-th/9803002
\bibitem{14}S.-J. Rey, J. Yee, hep-th/9803001
\bibitem{15}N. Drukker, D.J. Gross, H. Ooguri, Phys. Rev. D60 (1999) 125006, hep-th/9904191
\bibitem{16}H. Dorn, Fortschr. Phys. 34 (1986) 11
\end{thebibliography}
\end{document}